# Comprehensive Dosimetric Verification and Positional Sensitivity Analysis in Brachytherapy: A Unified ESAPI Tool for HDR and LDR Treatments

J. A. Válgoma

*Department of Medical Physics, Hospital Universitario Basurto, Bilbao, Spain*

**Abstract**

This study presents the development and validation of an independent software tool based on the Varian Eclipse Scripting API (ESAPI) for multi-modal brachytherapy Quality Assurance (QA). The tool addresses GEC-ESTRO HDR protocols and LDR positional uncertainty analysis. Engineered in C#, the application interfaces with BrachyVision, Vitesse, and Variseed, enabling independent TG-43 dose calculations—comparing point and line source models—integrated with EQD2-based radiobiological summation.

In HDR cervical cancer, the tool successfully automated EMBRACE II protocol reporting, streamlining clinical workflows by combining dosimetric QA with predictive and prospective planning. For LDR prostate treatments, a stochastic simulation module quantified the impact of systematic (rigid-body) versus random seed displacements on target coverage ($D_{90\%}$) and Organs at Risk (OAR) safety ($D_{0.1cc}$).

Sensitivity analysis in LDR prostate implants was benchmarked using two clinical cases (prostate volumes 31 cc and 71.3 cc). LDR simulations revealed that systematic displacements (±2 mm) yielded significantly higher dosimetric deviations than stochastic movements. In the 31 cc case, systematic shifts resulted in a rectal ($D_{0.1cc}$) standard deviation (SD) of 24.3 Gy, whereas random displacements reduced this to 12.4 Gy. In the 71.3 cc case, random displacements resulted in a rectal $D_{0.1cc}$ SD of 7.6 Gy, confirming that smaller volumes exhibit heightened sensitivity to errors. Technical analysis demonstrated that the point source model overestimated bladder $D_{10\%}$ by 8% relative to the line source model.

Our findings confirm that systematic rigid-body shifts represent a greater clinical risk for OAR toxicity than stochastic migration. Integrating predictive sensitivity analysis into the clinical workflow significantly enhances patient safety through robust plan verification.

# 1. Introduction

Modern brachytherapy has evolved significantly towards image-guided adaptive techniques (IGABT), a shift that has fundamentally improved clinical outcomes in both High-Dose-Rate (HDR) and Low-Dose-Rate (LDR) treatments[1]. However, this progress brings new complexities: the management of multi-modal data—specifically the integration of External Beam Radiation Therapy (EBRT) with brachytherapy boosts—and the inherent uncertainties in source positioning remains a formidable challenge in daily clinical practice.

In gynecological HDR brachytherapy, international protocols such as GEC-ESTRO and the EMBRACE II study[2] have standardized the reporting of dose-volume histogram (DVH) parameters. These standards emphasize the criticality of $D_{2cc}$ for organs at risk (OARs) and $D_{90\%}$, $D_{98\%}$ for target volumes. Despite the existence of these benchmarks, the extraction of radiobiological parameters like the Equivalent Dose in 2 Gy fractions (EQD2) often relies on manual data transcription into external spreadsheets. This fragmented workflow is not only time-consuming but also highly susceptible to human error, potentially compromising the accuracy of the total biological dose assessment. Furthermore, traditional planning tools typically lack integrated features for predictive dose summation, making it difficult to perform adaptive planning to compensate for dosimetric deviations from previous fractions. This lack of a unified, forward-looking system complicates the accurate assessment and optimization of the total biological dose delivered throughout the treatment course.

Similarly, LDR prostate brachytherapy with permanent $^{125}$I seed implants[3] faces the persistent challenge of post-implant source displacement[4,5]. This uncertainty often stems from the needle withdrawal process, where the friction between the needle and the seeds (or the surrounding tissue) can cause the seeds to be dragged or retracted from their intended positions. Furthermore, subsequent seed migration poses a long-term risk to plan stability. Clinical evidence suggests that source displacement can drastically reduce the Tumor Control Probability (TCP) and elevate toxicity levels in the urethra, bladder and rectum. While conventional Treatment Planning Systems (TPS) provide high-precision static dosimetric snapshots[6], they typically lack integrated modules to perform stochastic sensitivity analysis. Consequently, medical physicists often cannot evaluate the robustness of a plan against the dynamic reality of seed movement within the clinical software environment.

Building upon our previous research regarding ESAPI-based automation for EBRT (Válgoma et al., 2021)[7], this study presents a comprehensive, multi-modal software tool developed within the Eclipse Scripting API environment. This tool unifies the independent verification of HDR and LDR plans, ensuring strict adherence to international protocols[2,6,8] while introducing a novel simulation framework. For LDR treatments, the script enables a comparative analysis between point and line source formalisms[9,10] (TG-43) and performs Monte Carlo-like simulations to quantify the dosimetric impact of systematic versus random seed displacements.

The aim of this paper is to describe the technical implementation of this unified tool to analyze HDR gynecological cases and investigate how distinct displacement patterns and prostate volumes influence final LDR prostate dosimetry. By providing a new framework for "stress-testing" treatment plans, this tool offers a pathway toward enhanced patient safety and more robust brachytherapy planning.

## 2. Materials and Methods

### 2.1. Software Architecture and Development

The application was developed in C# using the .NET Framework and is integrated directly into the Varian Eclipse environment via the Eclipse Scripting API (ESAPI). To ensure a versatile clinical workflow, the tool was designed with a multi-modal interface capable of identifying and processing plans from various Varian planning systems, specifically BrachyVision and Vitesse for HDR treatments, and Variseed for LDR permanent implants. The graphical user interface (GUI) was engineered using Windows Presentation Foundation (WPF), allowing for seamless data visualization and integration within the existing clinical infrastructure.

## 2.2. HDR Brachytherapy Module and Quality Assurance

For HDR treatments our institution utilizes the Varian Bravos™ afterloader since 2020 after more than 30 years of experience with other solutions. The script was programmed to automatically extract critical plan parameters, including source dwell positions, dwell times, and 3D catheter reconstruction data. Three primary functionalities were implemented for HDR QA:

- Independent Dose Verification: An independent calculation engine based on the TG-43 formalism[6] was developed. This allows for point-dose, linear-dose or dose-matrix verification (including the ability to verify dose points within individual organs) to validate the primary results generated by BrachyVision or Vitesse. By performing these redundant calculations separately for target volumes and organs at risk, the tool provides an additional layer of verification and dosimetric integrity, ensuring the accuracy of the treatment plan before delivery.
- Radiobiological Summation: For gynecological cases (cervix), the tool adheres to the EMBRACE I/II (GEC-ESTRO) protocols[2]. It automatically calculates the Biological Effective Dose (BED) and the Equivalent Dose in 2 Gy fractions (EQD2) for both target volumes (HR-CTV, IR-CTV, GTV-RES) and organs at risk (rectum, bladder, sigmoid). These calculations assume $\alpha/\beta$ ratios of 10 Gy for tumor volumes and 3 Gy for healthy tissues. While the script utilizes the protocol-recommended default $\alpha/\beta$ ratios for tumor volumes and OAR, it is designed with a dynamic recalculation interface. This allows the user to manually modify these parameters and instantaneously update the biological dose values, providing a flexible framework for personalized radiobiological analysis or research-specific investigations.
- Predictive simulations: Beyond retrospective QA, the tool incorporates a predictive simulation module for gynecological treatments. It enables the summation of multiple treatment fractions—including future hypothetical fractions—while planning the initial stages. This allows clinical teams to project the cumulative dose to target volumes and OARs early in the treatment course, facilitating proactive plan adjustments. Furthermore, the script supports the simulation of additional treatment sessions to compensatively "rescue" the dosimetry in cases where early fractions did not meet clinical objectives. This feature provides a robust framework for adaptive brachytherapy, ensuring that the total delivered EQD2 remains optimized despite inter-fraction variations.

## 2.3. LDR Prostate Module and Displacement Simulation

The LDR module interfaces with Variseed to retrieve the exact planned coordinates (x, y, z) and orientation vectors of each implanted $^{125}$I seed[10]. To account for clinical dosimetric uncertainties, the following features were incorporated:

- Point vs. Line Source Comparison: The script performs dose calculations using both point and line source approximations[8,9,10]. This comparison is used to evaluate the impact of seed anisotropy, which is particularly significant in the steep dose gradients of near-source regions.
- Stochastic Displacement Simulator: A Monte Carlo-like simulation module was developed to model post-implant seed migration. The user can define a maximum displacement amplitude (e.g., ±3 mm in one or more directions) and execute three distinct scenarios:
1. Systematic (Rigid-body) Shift: All seeds are moved simultaneously in a uniform direction to simulate "in-block" migration, a phenomenon that could be observed in stranded seed configurations.
2. Random Displacement: Each seed is moved independently in a random direction to simulate loose seed migration or stochastic positioning errors.

3. Longitudinal Migration (Y-axis): A specific analysis focusing on movement along the needle insertion axis (Y-axis) is provided to evaluate the specific impact of cranial-caudal shifts on urethral, bladder and rectal dosimetry.

## 2.4. Clinical Benchmarking

For the HDR module, a retrospective cohort of 70 gynecological patients from our last 6-years of clinical practice was utilized to validate the automated GEC-ESTRO reporting and radiobiological summation. The analysis focused on the variation and stability of key dosimetric indices:

- Target Coverage: $D_{90\%}$ and $D_{98\%}$ for the HR-CTV, IR-CTV, GTV-RES.
- OAR Safety: $D_{2cc}$ for the rectum, bladder, and sigmoid colon.

For the LDR prostate module, two representative clinical cases with distinct volumes (31 c.c. and 71.3 c.c.) were selected to benchmark the stochastic displacement simulator. These cases were used to quantify how $^{125}I$ point and line source models affect the dosimetry and how prostate size influences the sensitivity of dose-volume histogram (DVH) parameters ($D_{90\%}$, $D_{0.1cc}$) to source migration. By comparing the script-generated results against the primary values (baseline) from the planning TPS (Variseed) we established the tool's reliability for high-precision clinical audits.

# 3. Results

## 3.1. Implementation and Automated Interface Analysis

The developed tool features a context-aware graphical user interface that automatically detects the type of brachytherapy plan (HDR or LDR) upon execution within the Eclipse environment (Figure 1a, Figure 1b). This automation eliminates the need for manual module selection, streamlining the clinical workflow. Once the plan type is identified, the interface dynamically populates the relevant analytical panels:

- For HDR plans: The tool performs an independent physical QA of the treatment plan for a line source model (Figure 2), verifying critical parameters such as source strength, dwell times, and TG-43 dose calculations. Specifically for cervical cancer cases, the interface additionally triggers the radiobiological summation module (BED/EQD2), extracting and calculating fraction-specific data according to GEC-ESTRO/EMBRACE protocols (see Figure 1b for a detailed explanation).
- For LDR plans: The interface activates the TG-43 verification for a point and line source model and stochastic sensitivity analysis modules (Figure 3).

In both modalities, the GUI provides immediate visual feedback via a "traffic light" quality assurance (QA) system based on institutional dose constraints. This allows the clinical team to instantaneously identify potential protocol deviations through a color-coded status (green for compliant, orange for warning threshold, red for exceeding limits).

**Figure 1a**: Screenshot of the main GUI demonstrating the automated plan detection of LDR Permanent I-125 Implant plan previously imported in ARIA for QA feedback.

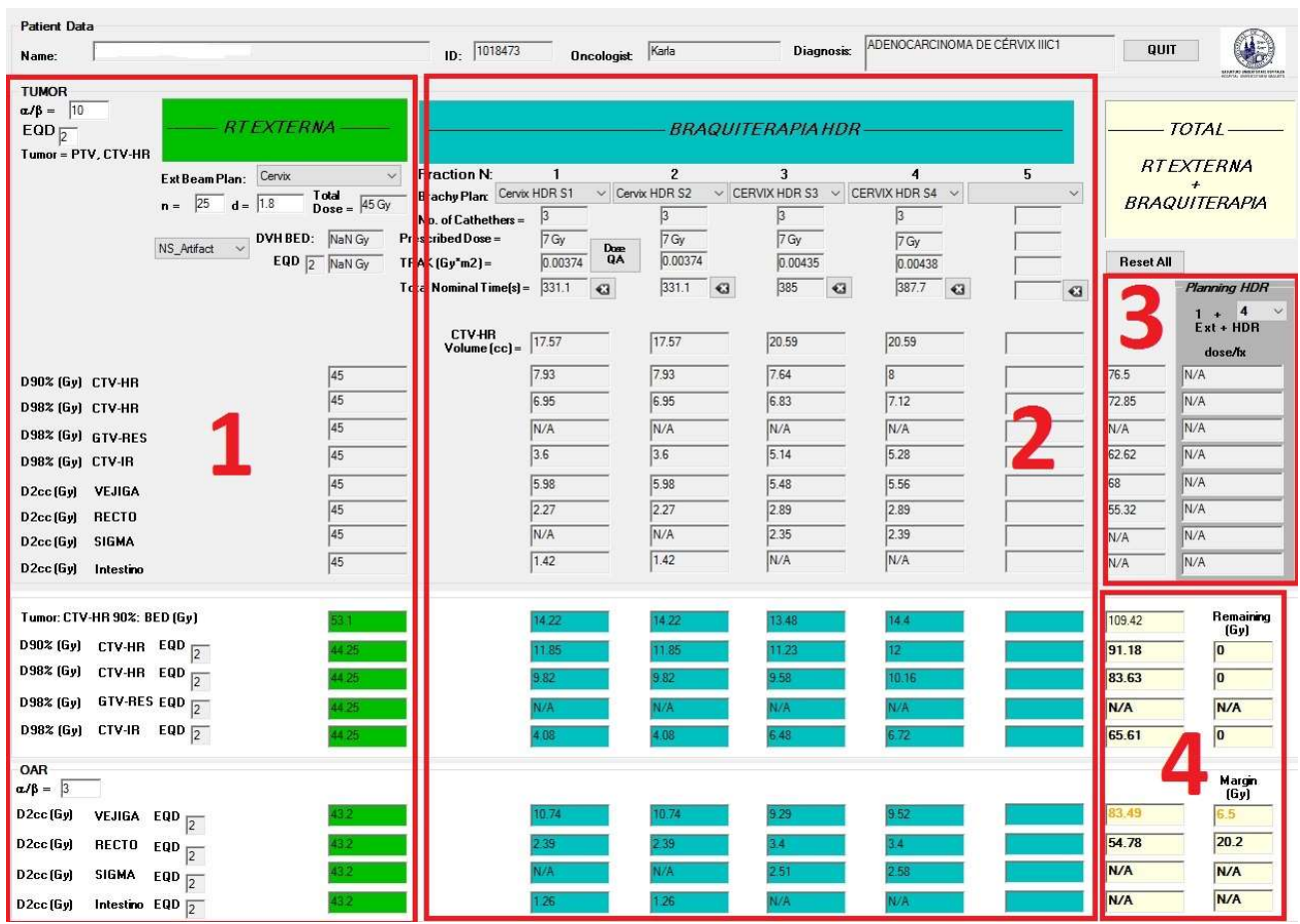

**Figure 1b**: Clinical Interface demonstrating the automated plan detection of a cervical HDR Brachytherapy plan for summation and prospective planning of EQD2 according to GEC-ESTRO/EMBRACE protocols for a sample patient. The software integrates External Beam Radiotherapy data (**1**) with individual HDR Brachytherapy fractions (**2**) calculating EQD2 for both Target Volumes and OAR. A key feature is the predictive planning module (**3**), which calculates the remaining dose capacity for upcoming fractions. Finally, Target Coverage and OAR Safety Margins are indicated via a "traffic light" quality assurance (QA) system based on institutional dose constraints (**4**).

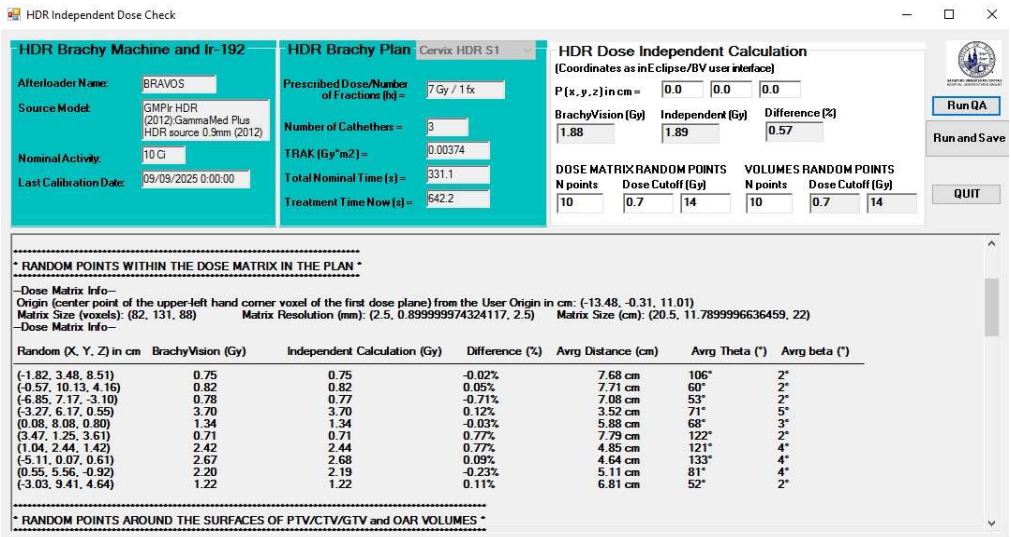

**Figure 2.** GUI for the independent dose verification of High-Dose Rate (HDR) brachytherapy plans. The interface displays Ir-192 source and afterloader model (left panel), alongside specific treatment plan parameters such as TRAK and total dwell time (center panel). The tool features interactive text boxes that allow the user to customize the number of evaluation points and dose thresholds for both the dose matrix and the target volumes and OAR. The bottom section provides a detailed point-by-point comparison between the Treatment Planning System (BrachyVision, Vitesse) and the independent TG-43 calculation for a HDR line source model, ensuring dosimetric accuracy before treatment delivery.

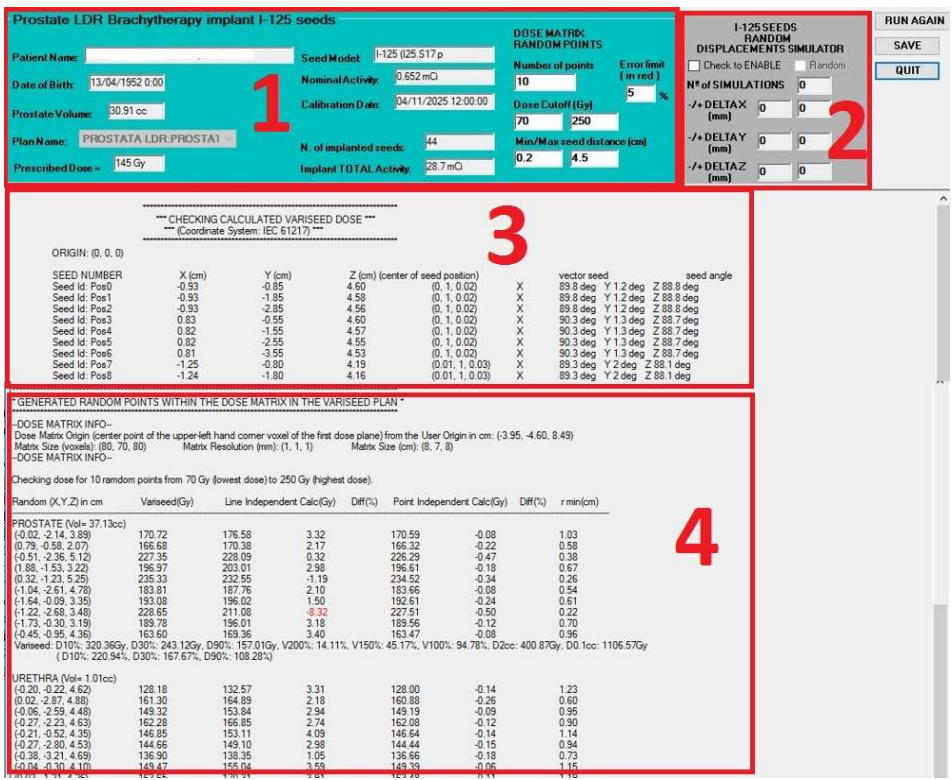

**Figure 3.** GUI for the independent quality assurance and sensitivity analysis of an I-125 LDR prostate brachytherapy plan (Variseed™). The interface displays patient-specific information and treatment parameters, including prostate volume, prescribed dose, seed model, nominal activity, and the total number of implanted sources (**1**), an advanced module for robustness analysis that allows the simulation of Systematic vs. Random seed movements where user can define displacement vectors and the number of simulations to evaluate the plan's sensitivity to potential source migration (**2**), the spatial coordinates and orientation for every seed in the implant (**3**), the comparison of the calculated dose by the TPS against independent TG-43 Line Source and Point Source models (**4**).

## 3.2. HDR Validation: Cervix and GEC-ESTRO Protocols

Validation was conducted on a cohort of 70 gynecological patients treated with MRI-guided brachytherapy and the Varian Bravos™ afterloader. The script achieved 100% accuracy in data extraction compared to manual entry. By automating the radiobiological summation (BED and EQD2), the tool reduced the reporting time from approximately 15 minutes to less than 1 minute per fraction. Furthermore, the integration of the ESAPI-based framework facilitates a prospective dosimetric assessment by projecting cumulative dose distributions from the initial fraction. This predictive capability transitions the workflow toward a pro-active adaptive planning strategy, allowing clinicians to anticipate cumulative $D_{90\%}$ and $D_{2cc}$ trends. Consequently, the tool provides a robust platform to implement inter-fraction compensatory adjustments, effectively mitigating the dosimetric impact of sub-optimal applicator geometry or anatomical variations encountered in subsequent fractions.

The tool allowed for the tracking of a significant institutional learning curve over a six-year period, as shown in the table below:

| Volume | Mean Dose (EQD2) | Initial Implementation Year | Six-Year Sustained Outcome |
|---|---|---|---|
| **TARGET** | $D_{90\%}$ HR-CTV | 86.2 Gy | 92.8 Gy |
| | $D_{98\%}$ HR-CTV | 76.0 Gy | 83.8 Gy |
| | $D_{98\%}$ RES-GTV | 89.6 Gy | 131.2 Gy |
| | $D_{98\%}$ IR-CTV | 59.9 Gy | 68.5 Gy |
| **OAR** | $D_{2cc}$ Bladder | 87.1 Gy | 75.7 Gy |
| | $D_{2cc}$ Rectum | 72.0 Gy | 56.3 Gy |
| | $D_{2cc}$ Sigmoid | 75.4 Gy | 53.3 Gy |

***Table1:*** *This table compares the mean dosimetric parameters of the patient cohort during the first year of implementation versus the outcomes achieved in the sixth year. The observed improvement in target coverage—specifically the significant escalation in $D_{98\%}$ for the RES-GTV—is attributed to a synergistic combination of factors: the support provided by the ESAPI script in ensuring dosimetric adherence, and the increased clinical experience of the radiation oncologists in applicator placement and MRI-guided adaptive strategies.*

This data confirms that the use of MRI-guided adaptive brachytherapy significantly improved target dose coverage while maintaining organ-at-risk safety.

## 3.3. LDR Sensitivity Analysis: Impact of Seed Displacement

The sensitivity analysis focused on the stochastic simulation of seed migration for two distinct prostate volumes (31 c.c. and 71.3 c.c.) using a $^{125}$I line source model.

A. **Systematic vs. Random Movement**: For the 31 c.c. prostate, a systematic shift of ±2 mm resulted in a rectal $D_{0.1cc}$ standard deviation (SD) of 24.3 Gy, whereas random movement reduced this uncertainty to 12.3 Gy and a bladder $D_{0.1cc}$ SD of 16.8 Gy, whereas random movement reduced this to 7.1 Gy. As illustrated in the sensitivity curves (Figure 4a, 4b), the results underscore that the rectum and bladder are highly sensitive to the 'block movement' of the implant. While stochastic seed migration (Random, solid lines) tends to average out the dose distribution—maintaining a relatively stable dose profile across all displacements—a systematic shift (dashed lines) induces high-amplitude dose fluctuations. Specifically, the dashed orange line representing systematic $D_{0.1cc}$ shows sharp dose spikes that are attenuated in the Random model. This indicates that a global translation of the seeds—such as that caused by needle withdrawal or global edema resolution—can translate the high-dose core directly into the rectal and bladder wall, drastically increasing the risk of radiation-induced proctitis and cystitis compared to independent seed migration.

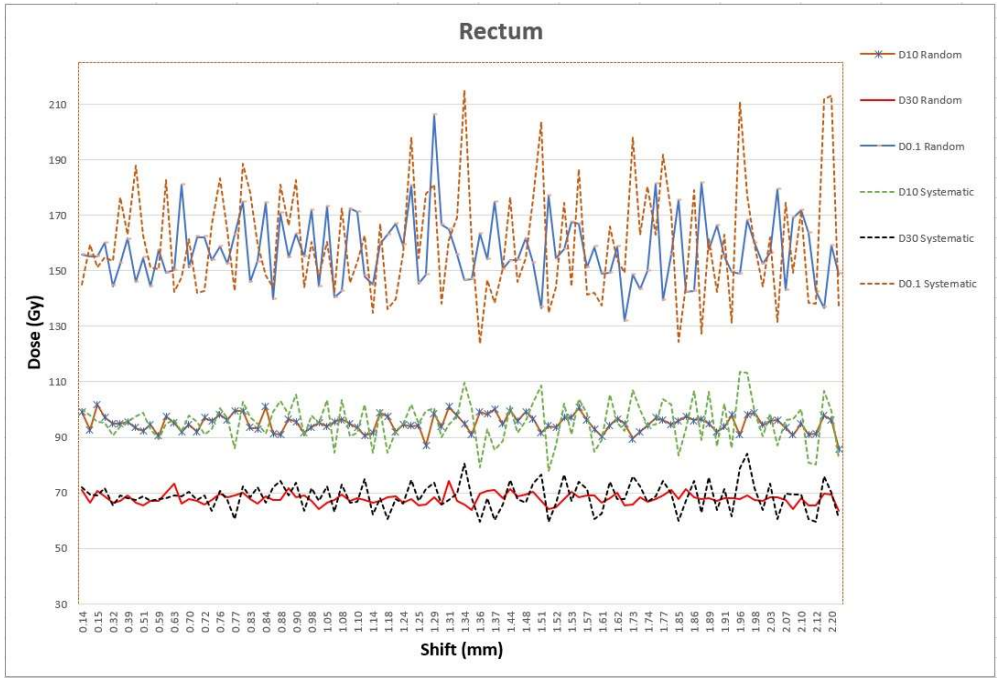

**Figure 4a.** Sensitivity analysis of rectum dosimetry in a 31 c.c. prostate patient as a function of seed displacement. The X-axis represents the Shift (the magnitude of the displacement vector applied to the seeds in mm), and the Y-axis represents the absorbed Dose in Gy. Solid lines denote random independent displacements, while dashed lines represent systematic block movements. Note the high volatility and dose spikes in the systematic $D_{0.1cc}$ (dashed orange line)

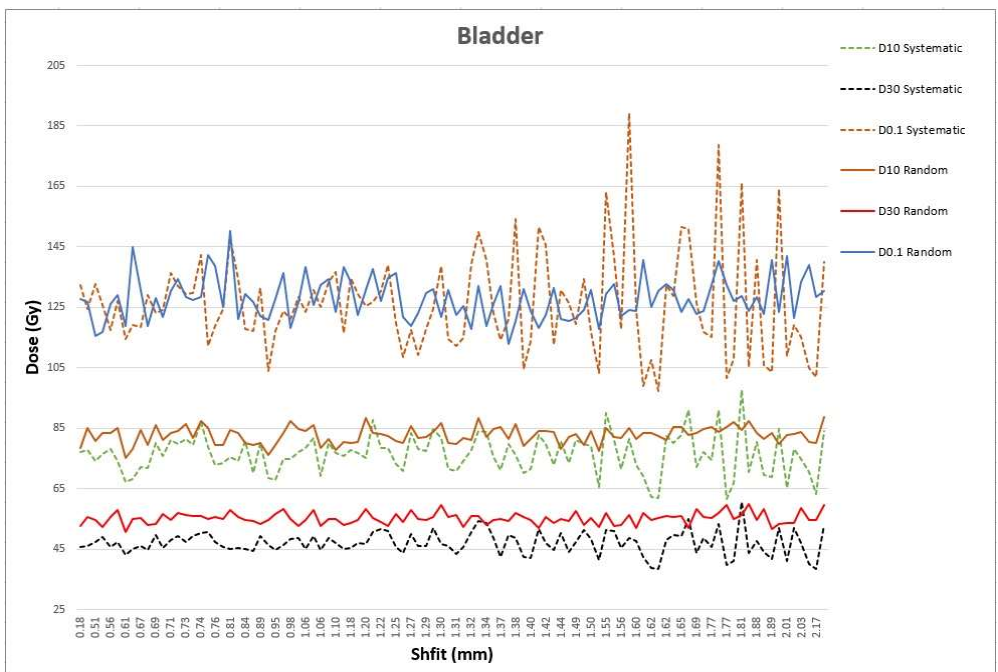

**Figure 4b.** Sensitivity analysis of bladder dosimetry in a 31 c.c. prostate patient as a function of seed displacement. Note the high volatility and dose spikes in the systematic $D_{0.1cc}$ (dashed orange line) compared to the stability of the random model, illustrating the increased risk of bladder toxicity associated with global implant translations.

B. **Volume Influence**: Dosimetric variability was found to be volume-dependent. The larger prostate (71.3 c.c.) exhibited a more stabilized response to random movements with a rectal $D_{0.1cc}$ SD of 7.6 Gy, compared to the smaller prostate with a rectal $D_{0.1cc}$ SD of 12.4 Gy (Figure 5) where individual seed shifts had a more pronounced local impact due to the proximity of the rectal wall. A 5 mm systematic shift in a 31 c.c. prostate plan can increase the bladder $D_{0.1cc}$ by over 235%, whereas a random shift of the same magnitude only results in a 1.5% variation. This variation is less important for larger prostate volumes (Table 2) that confirms that plan robustness is primarily threatened by collective displacements with a prostate volume dependence.

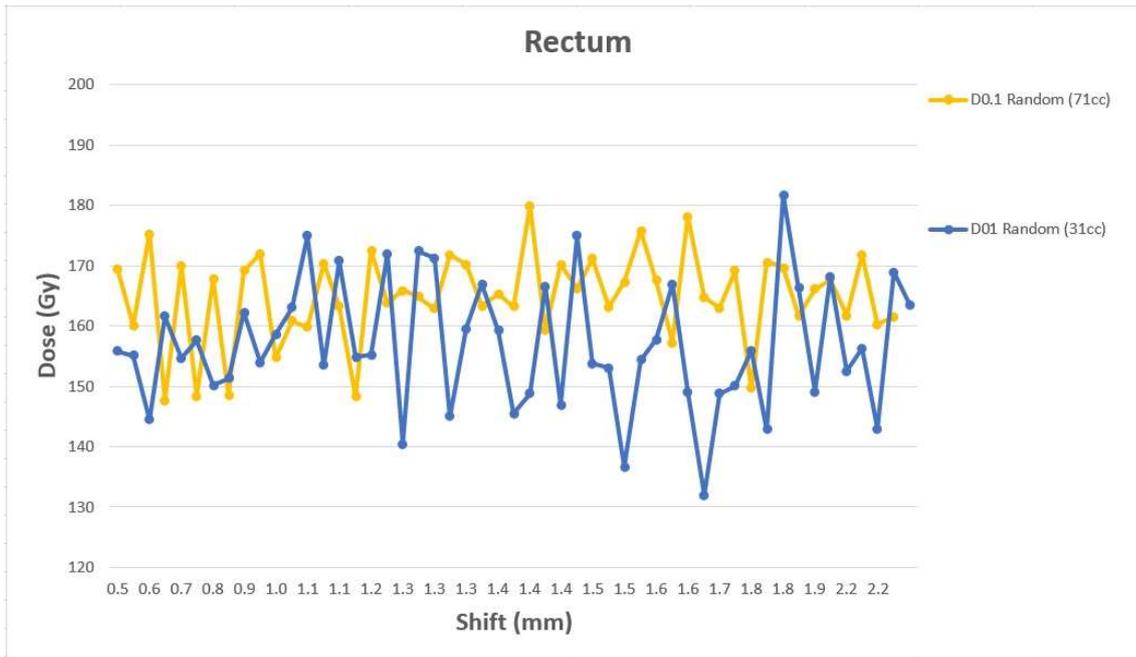

**Figure 5.** Impact of prostate volume on dosimetric stability for the rectum under random seed displacement. The graph compares the $D_{0.1cc}$ response for a large prostate (71.3 c.c., yellow line) and a small prostate (31.1 c.c., blue line). The X-axis represents the magnitude of the random Shift (mm) and the Y-axis the absorbed Dose (Gy). The larger volume exhibits a more stabilized response (SD = 7.6 Gy), whereas the smaller volume shows significantly higher variability (SD = 12.4 Gy). This demonstrates that in smaller prostates, individual seed shifts have a more pronounced local impact due to the steep dose gradients and the immediate proximity of the rectal wall.

| Volumen | Movement Type | Shift (mm) | Prostate D90% (Gy) | Var (%) | Bladder D0.1cc (Gy) | Var (%) | Rectum D0.1cc (Gy) | Var (%) | Urethra D0.1cc (Gy) | Var (%) |
|---|---|---|---|---|---|---|---|---|---|---|
| V= 31 cc | Baseline | 0 | 166,37 | - | 125,16 | - | 149,16 | - | 208,9 | - |
|  | Systematic | -5 | 125,4 | -24,63 | 69,7 | -44,31 | 293,87 | 97,02 | 198,05 | -5,19 |
|  | Random | -5 | 164,39 | -1,19 | 122,21 | -2,36 | 150,72 | 1,05 | 211,06 | 1,03 |
|  | Systematic | +5 | 158,75 | -4,58 | 420,23 | **235,75** | 121,84 | -18,32 | 212,46 | 1,70 |
|  | Random | +5 | 162,49 | -2,33 | 123,25 | **-1,53** | 172,58 | 15,70 | 207,73 | -0,56 |
| Volumen | Movement Type | Shift (mm) | Prostate D90% (Gy) | Var (%) | Bladder D0.1cc (Gy) | Var (%) | Rectum D0.1cc (Gy) | Var (%) | Urethra D0.1cc (Gy) | Var (%) |
| V=71.3 cc | Baseline | 0 | 170,83 | - | 169,82 | - | 174,73 | - | 201,86 | - |
|  | Systematic | -5 | 158,73 | -7,08 | 85,97 | -49,38 | 200,59 | 14,80 | 202,41 | 0,27 |
|  | Random | -5 | 174,57 | 2,19 | 238,74 | 40,58 | 161,66 | -7,48 | 203,8 | 0,96 |
|  | Systematic | +5 | 148,44 | -13,11 | 325,02 | **91,39** | 150,83 | -13,68 | 195,77 | -3,02 |
|  | Random | +5 | 167,12 | -2,17 | 244,04 | **43,71** | 159,58 | -8,67 | 201 | -0,43 |

*Table 2.* Dosimetric variation (Var %) following systematic and random seed displacements for two different prostate volumes. For systematic movements, a positive shift indicates a cranial displacement (toward the bladder), while a negative shift indicates a caudal displacement (toward the apex). Percent variations (Var %) are calculated relative to the static baseline plan.

## 3.4. Formalism Comparison: Point vs. Line Source

Previous results were obtained using a line source model for $^{125}$I seeds. However, the tool enables a direct comparison between TG-43 point and line source models. In near-source regions, the point source model consistently overestimate the dose in proximal OARs in accordance with literature[9] for our IsoSeed I25.S17plus $^{125}$I seed[11]. Specifically, for the 31 c.c. prostate volume patient, the point source model resulted in 82.5 Gy for bladder $D_{10\%}$ compared to 76.3 Gy calculated with the line source model—an 8% difference that could lead to overly conservative clinical decisions or misinterpretation of toxicity risks.

# 4. Discussion

## 4.1. Clinical Efficiency and Protocol Adherence

The integration of our ESAPI-based tool into the clinical workflow has fundamentally transformed the quality assurance (QA) process for HDR treatments. The reduction in reporting time from 15 minutes to less than 1 minute per fraction is not merely a gain in efficiency; it represents a significant reduction in the potential for human error associated with manual data transcription into external spreadsheets. By automating the radiobiological summation of EQD2 according to GEC-ESTRO/EMBRACE II guidelines, the tool ensures that every patient benefits from a rigorous, standardized dosimetric audit.

Beyond efficiency, the inclusion of a predictive dose summation module allows the clinical team to transcend retrospective analysis. By projecting cumulative doses from the very first fraction, physicists can identify potential OAR overdoses or target underdosages before the treatment course is completed. This "early warning" system enables compensatory adaptive planning, where subsequent fractions can be specifically optimized to "rescue" the total dosimetry—for instance, by escalating the dose in a final session to compensate for a sub-optimal applicator placement in an earlier one. This proactive approach is reflected in our institutional six-year learning curve, where target coverage was significantly improved without exceeding safety constraints.

## 4.2. Systematic vs. Random Migration: The "Dose Shift" Hazard

Our simulations highlight a critical distinction in LDR dosimetry regarding the nature of seed movement. Stochastic (random) migration often leads to "dose blurring," where individual errors partially cancel each other out, resulting in minimal deviations from the original plan. However, systematic (rigid-body) shifts—a known risk in clinical practice—cause a "dose shift" that can drastically alter the dose-volume histogram (DVH) of critical structures.

## 4.3. The Volume-Effect and Plan Robustness

The data in Figure 4a and Figure 5 reveals that prostate volume is a primary determinant of dosimetric stability. Smaller glands (31 c.c.) exhibit a much narrower safety margin compared to larger volumes (71.3 c.c.). In small prostates, the proximity of the seeds to the rectal wall means that even a minor systematic error results in an increase in dose; this is evidenced by the rectal $D_{0.1cc}$ standard deviation being nearly double in the 31 c.c. case (24.3 Gy vs. 12.4 Gy).

Larger glands benefit from a "statistical cushion"—a higher density of sources allows for a prompter averaging of positional uncertainties. This suggests that patients with smaller prostate volumes should be prioritized for the virtual "stress-testing" provided by our simulator.

### 4.4. Focal Boost Strategies and Plan Robustness

One of the most promising applications of this framework is the evaluation of focal dose escalation. Preliminary analysis suggests that achieving targeted dose intensification for dominant intraprostatic lesions is technically feasible through optimized seed spatial distribution. This approach could potentially allow for a more efficient use of sources, potentially requiring fewer needles and seeds.

The sensitivity module of our script demonstrates that such escalated plans can maintain a stability index comparable to standard whole-gland implants. This finding suggests that dose intensification does not inherently compromise plan robustness provided that the high-dose regions are strategically positioned and validated through the predictive sensitivity analysis developed in this work. This evidence establishes a clear path for future research into safer and more personalized focal treatments.

### 4.5. Technical Considerations: Point vs. Line Source

Finally, the 8% overestimation of bladder dose by the point source model compared to the line source model (TG-43) is a critical technical finding. Some institutions opt for the point source model as a way to simplify the planning process, under the assumption that this model's isotropy roughly accounts for the unpredictable movement and random tilting of seeds post-implant. However, our results suggest that this 'compensatory' simplification may be misleading.

In the near-source regions where OARs are often located, anisotropy matters and the dose distribution is highly sensitive to the geometric formalism. Relying on simplified point source models may lead clinicians to be overly conservative, potentially under-treating the target to "protect" an OAR based on inaccurate, higher dose readings.

# 5. Conclusions

The integration of a unified ESAPI-based tool for HDR and LDR brachytherapy has demonstrated significant advantages in terms of clinical efficiency and dosimetric safety. By automating data extraction and the calculation of radiobiological parameters, such as BED and EQD2, the tool effectively eliminates manual transcription errors and ensures strict adherence to international standards, including the GEC-ESTRO EMBRACE protocols in gynecological treatments. A key innovation of this software is its capability for predictive dose summation and adaptive planning. By allowing the clinical team to simulate future fractions and project cumulative doses early in the treatment course, the tool enables proactive plan modifications. This functionality allows for the "dosimetric rescue" of a treatment course by adjusting subsequent fractions to compensate for initial deviations, fundamentally shifting the clinical workflow from a passive retrospective audit to an active inter-fraction strategy.

The most critical finding of this study is the quantified distinction between systematic and random source displacements in LDR prostate treatments. Our results indicate that systematic rigid-body shifts—often associated with stranded seed configurations—pose a substantially higher risk to organs at risk (OARs) than stochastic migration. The high dosimetric variability observed in the rectum and bladder during systematic shifts underscores the necessity of performing sensitivity analysis as a standard component of the routine quality assurance workflow.

Furthermore, the comparative analysis between point and line source formalisms confirms that the point source model tends to overestimate doses in proximal critical structures. This effect is particularly pronounced in small-volume prostates where local dose gradients are steeper and safety margins are narrower. The ability to simulate these clinical scenarios within the Eclipse environment using a customized script provides medical physicists with a powerful "stress-test" for treatment plans before they are executed.

In conclusion, the developed application represents a robust and versatile solution for independent dose verification. It successfully bridges the gap between static treatment planning and the dynamic reality of post-implant source behavior. Theoretically, this framework could allow for the exploration of higher target doses to improve local control while maintaining strict adherence to healthy tissue safety thresholds. Ultimately, this ESAPI-based tool serves as a critical bridge toward a more proactive and adaptive brachytherapy workflow, enhancing the precision of both HDR and LDR modalities.

**Declaration of Competing Interest**

The author have no competing financial interests or personal relationships that could have appeared to influence the work reported in this paper.